\documentclass[aps,prd,nofootinbib,twocolumn,superscriptaddress,floatfix]{revtex4}
\usepackage{mathrsfs}
\usepackage{latexsym}
\usepackage{amsmath}
\usepackage{amssymb}
\usepackage{graphicx}
\usepackage{longtable}
\usepackage{multirow}
\usepackage{bbm}
\usepackage{epsfig}
\usepackage[usenames]{color}
\usepackage[colorlinks=true,citecolor=blue,linkcolor=blue]{hyperref}
\allowdisplaybreaks

\begin{document}



\title{Stochastic gravitational waves produced by the first-order QCD phase transition}

\author{Xu Han}
\author{Guo-yun Shao}
\affiliation{School of Physics, Xi'an Jiaotong University, Xi'an, 710049, China}

\begin{abstract}
We investigate the stochastic gravitational waves background arising from the first-order QCD chiral phase transition, considering three distinct sources: bubble collisions, sound waves, and fluid turbulence. Within the framework of the  Polyakov-Nambu-Jona-Lasinio (PNJL) model, we calculate the parameters governing the intensity of the gravitational wave phase transition and determine their magnitudes along the adiabatic evolutionary path. We introduce the effective bag constant $B_{\mathrm{eff}}$ related to the dynamical evolution of quarks to evaluate the intensity of the phase transition. By calculating expanded potential at the point of false vacuum, we find that all the bubbles are in the mode of runaway, leading the velocity of the bubble wall to the speed of light. The resulting gravitational wave energy spectrum is estimated, revealing a characteristic amplitude of the generated gravitational waves within the centihertz frequency range. We present the gravitational wave spectrum and compare it with the sensitivity range of detectors, and find that the gravitational wave spectra generated by these sources have the potential to be detected by future detectors such as BBO and $\mu$ARES.
\end{abstract}


\maketitle

\section{introduction}

Gravitational waves (GWs), as predicted by general relativity, are generated by alterations in space-time curvature and propagate outwards at the speed of light in wave-like form. Nowadays, gravitational wave signals from point sources have been successfully detected multiple times, such as the generated GWs form the binary compact objects~\cite{LIGOScientific:2016aoc,LIGOScientific:2017vwq}. However, there has been no observational data for the stochastic gravitational wave background (SGWB). Until recently, NANOGrav released their data, claiming that it might be possible to detect the SWGB signals within a certain error range~\cite{NANOGrav:2023hde,NANOGrav:2023gor}, European PTA~\cite{Antoniadis:2023utw,Antoniadis:2023rey}, Parkes~\cite{Zic:2023gta,Reardon:2023gzh} and Chinese~\cite{Xu:2023wog} PTAs also reported their data, which has attracted ample attention in the astrophysics, cosmology and particle physics communities.

The primary source of the stochastic gravitational wave background is phase transitions that occurred in the early universe during its evolution. The early universe undergoes multiple phase transitions during its development. If these changes were of a strong first order, the scalar field shifts from a symmetric phase to a broken one through quantum tunneling or thermal fluctuations. This would result in bubble nucleation within the unstable sea. If this
process is sufficiently rapid compared to the expansion rate determined by the Hubble parameter $H$, these bubbles would expand and collide, generating gravitational waves~\cite{witten1984cosmic, hogan1986gravitational, Tamanini:2016zlh, Athron:2022mmm}. Due to the minimal interaction of the gravitational waves with matter, they overlay and
permeate the entire universe, giving rise to what is known as the Stochastic Gravitational
Wave Background (SGWB).

According to modern cosmology, approximately within $10^{-6}\,s$ after the Big Bang, the
universe was primarily composed of a quark-gluon plasma (QGP) made up of free quarks
and gluons. This state has what is known as chiral symmetry. With the expansion and
cooling of the universe, when the energy reached about $100\,\mathrm{MeV}$, quarks and gluons began to
bind together to form hadrons. This process is known as color confinement. Simultaneously,
chiral symmetry was broken, implying that left-handed and right-handed quarks are no
longer identical. This marked the occurrence of the QCD chiral phase transition. Although the standard model of particle physics does not discuss such phase transitions, they are predicted by non-perturbative QCD~\cite{Chu:2022mjw, Sagunski:2023ynd}.

The physical image of the entire process is particularly intriguing. The scalar field initially lies at the global minimum of its potential function. As the first order chiral phase transition happens, some of the parameters change, a new minimum appears, gradually replacing the original minimum as the parameter continues to change, triggering vacuum decay~\cite{Coleman:1980}. If the changing parameter is the background temperature, this decay is known as thermal decay, leading to bubble nucleation~\cite{Linde:1983}. After bubble nucleation, the pressure difference inside and outside the bubble drives the bubble wall’s expansion. Once bubbles expand beyond inter-bubble separation, they inevitably collide. These collisions release kinetic energy stored in the bubble walls, transforming it into gravitational waves carrying energy and momentum of the phase transition. Considering all other effects, in summary, the generation of the power spectrum of stochastic gravitational waves from FOPT is the combined result of three key mechanisms: bubble wall collisions triggered by scalar field dynamics~\cite{Kosowsky:1992vn, laine1994bubble}, sound waves produced by fluid motion~\cite{Hindmarsh:2013xza, Hindmarsh:2017gnf}, and MHD turbulence induced by plasma dynamics~\cite{Caprini:2006jb}. 

Our work focus on  the gravitational wave background generated by the  first-order QCD phase transition in the early universe in the scenario of little inflation~\cite{Boeckel:2010}. The calculation is performed within the framework of one of the effective models describing strong-coupling QCD, the PNJL model~\cite{Ratti:2005jh}. We adopt a reasonable assumption that the bubbles evolve adiabatically along the path $s/\rho_B = \mathrm{const}$~\cite{reiter1998entropy,gunther2017qcd,Shuve:2017jgj} from the local minimum to the true vacuum, providing relatively exact values for the energy budget parameter $\alpha$ along each evolution line. Furthermore, we use the method of replacing the local minimum with its second-order Taylor expansion to derive the criterion for whether the bubble expansion is runaway or not~\cite{Chen:2017cyc, Li:2021qer}. And we find that for bubbles evolving along the adiabatic approximation evolution line, not reaching runaway is an impossible situation. Then the gravitational wave signals corresponding to the three situations are detailed and compared with the detection capabilities of current gravitational wave detectors. The results show that some gravitational signals has strong potential to be detected in future observations.

This paper is divided into three comprehensive sections. In Sec.\ref{sec2}, we offer a brief introduction of the PNJL model, and use which to explicitly calculate the thermodynamic parameters of the first-order chiral phase transition for later computation. In Sec.\ref{sec3A}, we study the chiral phase transition of QCD, and provide a detailed description of the adiabatic evolution and the criteria for bubble runaway expansion, from which we obtain the key parameters used for gravitational wave signal calculations. In Sec.\ref{sec3B}, we calculate the gravitational wave spectrum, and further determine the corresponding GW signal and evaluate the prospects for detectability by future observatories. Finally, a summary is given in Sec.\ref{sec5}.

\section{Polyakov-Nambu-Jona-Lasinio model}
\label{sec2}
The PNJL model is widely used in the description of QCD phase transitions 
The Lagrangian density in the PNJL model is formulated as
\begin{eqnarray}
\mathcal{L}_{\mathrm{PNJL}} &=&\, \bar{q}\,(i\gamma^\mu D_\mu+\gamma_0\hat{\mu}-\hat{m_0})\,q \nonumber \\
&&+\,G\sum^8_{k =0}\left[(\bar{q}\lambda^kq)^2+(\bar{q}\,i\gamma_5\lambda^kq)^2\right] \nonumber \\
&&-\,K\left[\mathrm{det}_f(\bar{q}\,(1+\gamma_5)\,q)+\mathrm{det}_f(\bar{q}\,(1-\gamma_5)\,q)\right] \nonumber \\
&&-\,U(\Phi,\bar{\Phi};T).    
\label{eq1}    
\end{eqnarray}

The covariant derivative in the Lagrangian is defined as $D_\mu=\partial_\mu-iA_\mu$.
The gluon background field $A_\mu=\delta_\mu^0A_0$ is supposed to be homogeneous
and static, with  $A_0=g\mathcal{A}_0^\alpha \frac{\lambda^\alpha}{2}$, where
$\frac{\lambda^\alpha}{2}$ is $SU(3)$ color generators.
The effective potential $U(\Phi[A],\bar{\Phi}[A],T)$ is expressed with the traced Polyakov loop
$\Phi=(\mathrm{Tr}_c L)/N_C$ and its conjugate
$\bar{\Phi}=(\mathrm{Tr}_c L^\dag)/N_C$. The Polyakov loop $L$  is a matrix in color space
\begin{equation}
   L(\vec{x})=\mathcal{P} \exp\bigg[i\int_0^\beta d\tau A_4 (\vec{x},\tau)   \bigg],
\end{equation}
where $\beta=1/T$ is the inverse of temperature and $A_4=iA_0$.


In Eq.~(\ref{eq1}),  $U$ symbolizes the effective potential energy of the Polyakov loop, which is temperature-dependent. Its role is to encapsulate the kinetic energy term associated with gluons. However, its selection is not unique. By juxtaposing this with results from lattice QCD simulations, a logarithmic effective potential can be ascertained, which is given as follows~\cite{Schaefer:2009ui}
\begin{equation}
\begin{aligned}
\frac{U(\Phi,\bar{\Phi};T)}{T^4}=&\,b(T)\,\mathrm{ln}\left[1-6\bar{\Phi}\Phi+4(\bar{\Phi}^3+\Phi^3)-3(\bar{\Phi}\Phi)^2\right]\\
&-\frac{a(T)}{2}\bar{\Phi}\Phi,  
\end{aligned}
\end{equation}
where $a(T)=a_0+a_1\left({T_0}/{T}\right)+a_2\left({T_0}/{T}\right)^2$, $b(T)=b_3\left({T_0}/{T}\right)^3$.
The parameters in the model are determined through an analysis of QCD thermodynamics and by fitting to the pure canonical region of QCD grid points~\cite{Kaczmarek:2002mc,Kaczmarek:2007pb}, as listed in the Table. \ref{tab1}.
\begin{table}[h]
\centering
\caption{PNJL model parameters}
\begin{tabular}{ccccc}
\toprule
$a_0$ & $a_1$ & $a_2$ & $b_3$ & ~~$T_0/MeV$~~ \\
\hline
~~~~3.51~~~~ & ~~~~-2.47~~~~ & ~~~~15.2~~~~ & ~~~~-1.75~~~~ & ~~~~210~~~~ \\ 
\hline
\end{tabular}
\label{tab1}
\end{table}

According to the temperature field theory, the giant thermodynamic potential of the system is as follow
\begin{eqnarray}
\Omega_{PNJL}&=&-\,2\,T\sum_{i=u,d,s}\int\frac{d^3p}{({2\pi})^3}(F_1+F_2) \nonumber \\
&&-\,2\int_\Lambda\frac{d^3p}{({2\pi})^3}3\,(E_u+E_d+E_s)+2\,G\sum_{i=u,d,s}\phi_i^2 \nonumber \\
&&-\,4\,K\phi_u\phi_d\phi_s+U(\Phi,\bar{\Phi};T)+C,
\label{eq2}    
\end{eqnarray}
where $E_i=\sqrt{\textbf{\textit{p}}^2+M_i^2}$, $F_1=\mathrm{ln}(1+3\Phi e^{{-(E_i-\mu_i)}/{T}}+3\bar{\Phi}e^{{-2(E_i-\mu_i)}/ {T}}+e^{{-3(E_i-\mu_i)}/{T}})$ and $F_2=\mathrm{ln}(1+3\bar{\Phi} e^{{-(E_i+\mu_i)}/{T}}+3\Phi e^{{-2(E_i+\mu_i)}/{T}}+e^{{-3(E_i+\mu_i)}/{T}})$.

The thermodynamic potential involves five independent variables: $\phi_u$, $\phi_d$, $\phi_s$, $\Phi$ and $\bar{\Phi}$. By examining the extremal values of the thermodynamic potential with respect to these variables, their corresponding values can be determined. For the clarification of our following computation, we present the expression of the energy density
\begin{eqnarray}
\epsilon&=&\,T\,s+\sum_{i=u,d,s}\rho_i\,\mu_i+\Omega_{PNJL} \nonumber \\
&=&-\,2\,N_c\sum_{i=u,d,s}\int_\Lambda\frac{d^3p}{({2\pi})^3}E_i \nonumber \\
&&+\,6\sum_{i=u,d,s}\int\frac{d^3p}{({2\pi})^3}E_i\left[n(E_i-\mu_i)+\bar{n}(E_i+\mu_i)\right] \nonumber \\
&&+\,2\,G\sum_{i=u,d,s}\phi_i^2-4\,K\phi_u\phi_d\phi_s \nonumber \\
&&+\,U(\Phi,\bar{\Phi};T)-T~\frac{\partial U (\Phi,\bar{\Phi};T)}{\partial T}+C    
\end{eqnarray}

It is important to highlight that the PNJL model requires the determination of five parameters. These parameters are obtained through fitting: the values of the masses of the $\pi$, $\mathbf{K}$, $\eta^\prime$ mesons: $m_\pi = 135.0$ MeV, $m_K=497.7$ MeV, and $m_{\eta^\prime}=957.8$ MeV, as well as the $\pi$ meson decay constant $f_\pi=92.4$ MeV~\cite{Karsch:2001cy}. The specific values of these parameters for the subsequent calculations are as follows: the $u$ and $d$ quark masses $m_u = m_d = 5.5$ MeV, the strange quark mass $m_s = 140.7$ MeV, the coefficient for the four-fermion interaction $G\Lambda^2 = 1.835$, the coefficient for the six-fermion interaction $K\Lambda^5 = 12.56$, and the three-momentum cut-off $\Lambda = 602.3$ MeV. These parameter values remain unchanged throughout the subsequent calculations.

\section{Numerical results and discussions}
\label{sec3}

\subsection{Bubble evolution with a first-order phase transition in the PNJL model}
\label{sec3A}

The physical process of gravitational wave production by first-order phase transitions is described by bubble dynamics. The final shape of the gravitational wave power spectrum is mainly controlled by two key parameters, $\alpha$ and $\beta$. Fortunately, both can be determined by making reasonable assumptions. Regarding the energy budget $\alpha$, its magnitude depends on the difference between the field values at the global minimum and the local minimum~\cite{espinosa2010energy}, which is also defined as the ratio of the vacuum energy density released by the phase transition to the background radiation energy density
\begin{equation}
\alpha=\left.\dfrac{\Delta\epsilon_{\mathrm{vac}}}{\epsilon_{\mathrm{rad}}}\right|_{T=T_\star},
\end{equation}
where $T_\star$ is the transition temperature. $\epsilon_{rad}$ describes the energy contribution of the radiation part in the quark field
\begin{equation}
\epsilon_{\mathrm{rad}}=6\sum_{i=u,d,s}\int\frac{d^3p}{({2\pi})^3}E_i\left[n(E_i-\mu_i)+\bar{n }(E_i+\mu_i)\right].
\end{equation}
The energy budget has been studied in the bag model~\cite{Giese:2020rtr, Wang:2023jto} 
with
\begin{equation}
\Delta\epsilon_{\mathrm{vac}}=\epsilon -\epsilon_{\mathrm{rad}}=B,
\end{equation}
where $B$ is the bag constant.

However, the simple bag model is not sufficient to accurately describe the strength of the phase transition. Therefore, inspired by the Bag model, in this study we construct the effective bag constant with the dynamical quark mass in the PNJL model, which is defined as
\begin{equation}
B_{\mathrm{eff}}=\epsilon -\epsilon_{\mathrm{rad}},
\end{equation}
and the expression of the effective bag constant~\cite{Shao:2011ij} is 
\begin{eqnarray}
B_{\mathrm{eff}}&=&-\,2\,N_c\sum_{i=u,d,s}\int_\Lambda\frac{d^3p}{({2\pi})^3}E_i \nonumber \\
&&+\,2\,G\sum_{i=u,d,s}\phi_i^2-4\,K\phi_u\phi_d\phi_s \nonumber \\
&&+\,U(\Phi,\bar{\Phi};T)-T\,\frac{\partial U(\Phi,\bar{\Phi};T)}{\partial T}+C.    
\end{eqnarray}

We find that $B_{\mathrm{eff}}$ behaves similarly as the bag constant in the MIT-Bag model. However, its value depends not only on the interaction between different quarks, but also on the density and temperature. And with $B_{\mathrm{eff}}$, we can now rewrite the energy budget in the following form
\begin{equation}
\alpha=\left.\dfrac{\Delta B_{\mathrm{eff}}}{\epsilon_{\mathrm{rad}}}\right|_{T=T_\star}=\left.\dfrac{{B_{\mathrm{eff}}}_+-{B_{\mathrm{eff}}}_-}{{\epsilon_+-B_{\mathrm{eff}}}_+}\right|_{T=T_\star}.
\end{equation}
For the bubble, the subscript ``+'' denotes the state that has not yet decayed and remains in a metastable state, while the subscript ``--'' represents the state that has decayed and transitioned into a vacuum state.

In the early universe, the transition from the quark phase to the hadronic phase is generally along a path with a small $s/\rho_B$, corresponding to lower baryon chemical potential. However, the assumption of little inflation suggests a possibility of a first-order phase transition from a high baryon chemical potential path, dominated by the hot big bang scenario of cosmic evolution~\cite{Boeckel:2010}. Therefore, in this scenario it is necessary to consider the changes in baryon chemical potential when computing the energy budget of a first-order transition. 

\begin{figure}[htbp]
\begin{center}
\includegraphics[scale=0.35]{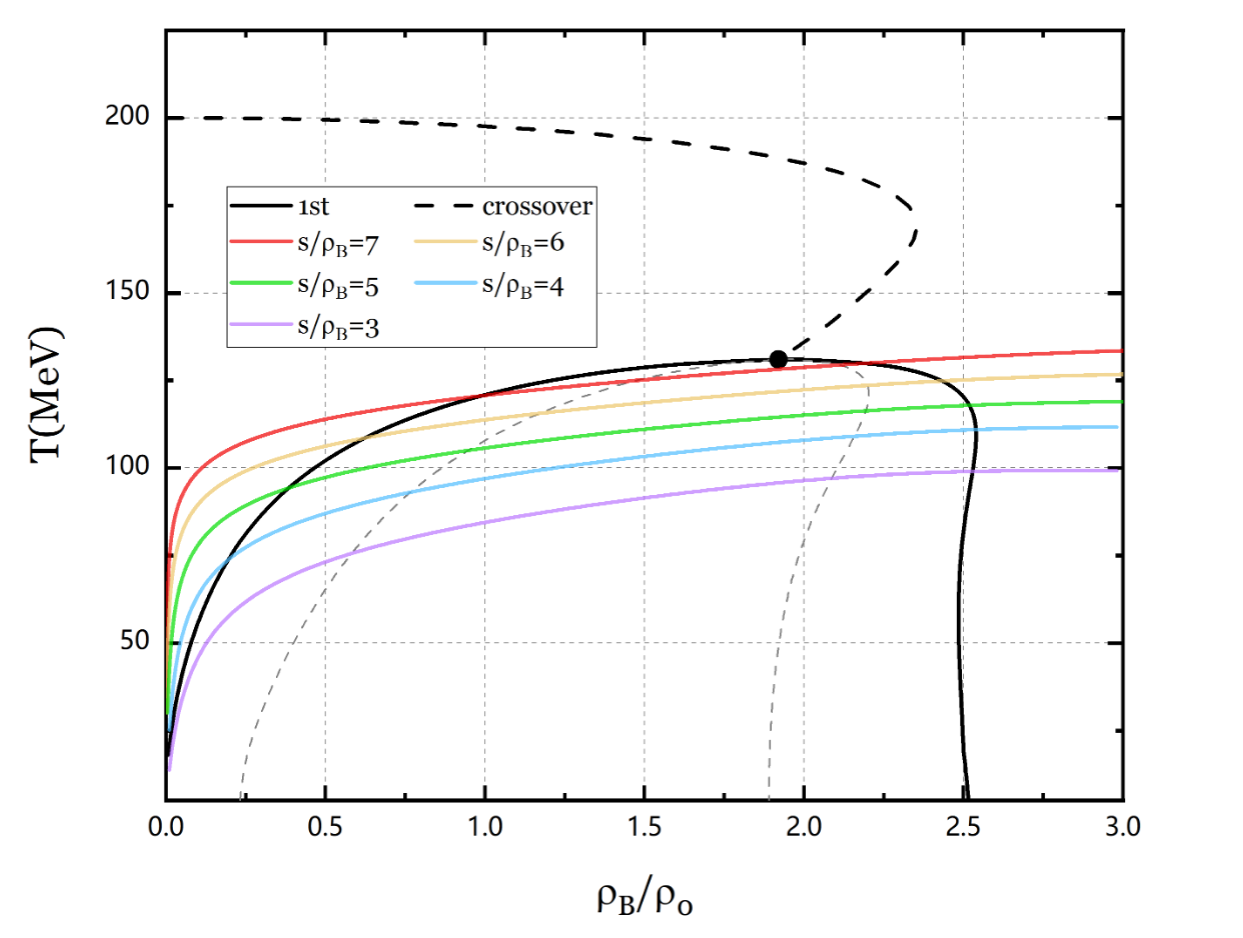}
\caption{The phase diagram illustrates the evolution paths of bubbles under different adiabatic conditions. The black solid line denotes the critical line of the QCD first-order chiral phase transition. The black dashed line indicates the crossover line of the chiral phase transition. Each colored solid line represents the adiabatic evolution curve corresponding to different values of $s/\rho_B$.}
\label{fig:s_rho}
\end{center}
\end{figure}

Since the local minima are distributed throughout the spatial grid,  the choice of the position of the false vacuum is somewhat arbitrary. Without specifying the evolution of the bubbles, the value of $\alpha$ cannot be determined but only approximately constrained. In this paper, we adopt a reasonable assumption that the bubbles evolve adiabatically along the path from the local minimum to the true vacuum, providing relatively exact values for $\alpha$ along each evolution line. This path is characterized by a constant value of $s/\rho_B$ as shown in Fig.~\ref{fig:s_rho}, where $s$ represents the entropy density and $\rho_B$  is the baryon number density.

The other important quantity is denoted as $\beta$, which characterizes the duration of the phase transition. It is defined as
\begin{equation}
    \beta = \left. H(T_\star)T_\star\cdot\frac{\mathrm{d}(S_3/T)}{\mathrm{d}T}\right|_{T=T_\star}
\end{equation}
Here, $S_3$ denotes the three-dimensional Euclidean bounce action. This parameter has been accurately given by solving the evolution variance numerically in many other outstanding works~\cite{Helmboldt:2019pan, Chen:2022cgj, Bai:2023cqj}. Here we refer to the value $\beta=4.75 \times 10^4$ from the literature~\cite{Helmboldt:2019pan} for calculations and discussions.

The speed of the bubble wall is another crutial parameter in the description of bubble evolution. After the bubbles form, their expansion mechanism depends on the resistance provided by the background fluid. Specifically, it can be divided into runaway and non-runaway situations, the former allowing the bubble wall to reach the speed of light due to insufficient resistance, while the latter implies the bubble wall which under a stronger resistance stabilizes at a lower speed, but is still highly likely to be relativistic. 

To study whether the bubble is in a runaway state or not, we use the method of replacing the local minimum with its second-order Taylor expansion to derive the criterion for whether the bubble expansion is runaway or not, if the value of the effective potential at the point of the true vacuum is lower than that at the false vacuum, then the bubble will runaway. We define an intensity factor $\alpha_{\infty}$ at infinity, which is represented as
\begin{equation}
\alpha_{\infty}=\frac{30}{24\pi^2}\frac{\sum_i\frac{1}{2}N\Delta m_i^2}{g_\star T_\star^2},
\end{equation}
where $c_i=n_i(c_i=n_i/2)$ represents the degrees of freedom for boson (fermion) particles, and $\Delta m_i^2$ is the square of the mass difference between the two phases. In the absence of a large-mass gauge boson, a condition of $\alpha>\alpha_\infty$ will result in an uncontrolled bubble expansion, enabling the bubble wall velocity to reach the speed of light and thus runaway~\cite{bodeker2009can}. At the runaway juncture, the energy contribution distribution index of various components can be defined by the following formula~\cite{Caprini:2015zlo}
\begin{eqnarray}
&\kappa_{\mathrm{en}}=1-\frac{\alpha_\infty}{\alpha},\\[5pt]
&\kappa_{\mathrm{sw}}=\frac{\alpha_\infty}{0.73+0.083\sqrt{\alpha_\infty}+\alpha_\infty},\\[5pt]
&\kappa_{\mathrm{tu}}=0.05\kappa_{\mathrm{sw}},
\end{eqnarray}
where $\kappa_{\mathrm{en}}$, $\kappa_{\mathrm{sw}}$ and $\kappa_{\mathrm{tu}}$ are the efficiency factors of bubble collision, sound waves and turbulence, respectively. Given that the bubble attains the speed of light, its wall velocity is $v_w=1$.

When $\alpha<\alpha_\infty$, the bubble will not runaway. Nevertheless, upon calculation through the PNJL model, it's found that for bubbles evolving along the adiabatic approximation evolution line, this inequality only holds when the phase transition intensity is extremely severe. In terms of the phase diagram, this means that only when the bubble evolves along a path with a lower value of $s/\rho_B$ will it not runaway. However, when the value of $s/\rho_B$ is small, the bubbles nucleated in the false vacuum won't reach the true vacuum along the evolution line, as both the temperature and the chemical potential will approach zero after evolution. Thus, from a physical point of view, a non-runaway scenario cannot be realized. Therefore, we specifically considers the situation where the bubbles runaway. In the subsequent calculation of the gravitational wave spectrum, we choose the case of $s/\rho_B=4$ for calculation.

\subsection{Stochastic gravitational waves}
\label{sec3B}

Bubble collisions release kinetic energy stored in the bubble walls, transforming it into gravitational
waves carrying energy and momentum of the phase transition. However, the sources of gravitational waves are not solely from bubble wall collisions but also originate from the background fluid itself. As bubbles expand within the background fluid, the movement of their walls stimulates fluid to form a persistent sound shell, converting some vacuum energy into kinetic energy, making sound waves an important source of gravitational waves. Meanwhile, the turbulent motion of charged particles in the plasma state of the background fluid causes fluctuations in the magnetic field, forming magnetohydrodynamic (MHD) turbulence, another significant source of gravitational waves~\cite{Hogan:1986}. Therefore, the generation of the gravitational wave power spectrum is a combined result of three key mechanisms: bubble wall collisions induced by scalar field dynamics, sound waves derived from fluid motion, and MHD turbulence induced by plasma dynamics. Therefore, the total energy spectrum of gravitational waves can be encapsulated in the following expression
\begin{equation}
h^2\Omega(f)=h^2\Omega_{\mathrm{en}}(f)+h^2\Omega_{\mathrm{sw}}(f)+h^2\Omega_{\mathrm{tu}}(f).
\end{equation}

The fitting formula for the gravitational wave energy spectrum, which is widely acknowledged in current studies due to comprehensive analyses and numerical simulations, can be represented as~\cite{huber2008gravitational,caprini2009stochastic,caprini2009general}
\begin{eqnarray}
&h^2\Omega_{\mathrm{en}}(f)=h^2{\Omega_{\mathrm{en,0}}}\frac{3.8\left(\frac{f}{f_{\mathrm{en}}}\right)^{2.8}}{1+2.8\left(\frac{ f}{f_{\mathrm{en}}}\right)^{3.8}},\\[5pt]
&h^2\Omega_{\mathrm{sw}}(f)=h^2{\Omega_{\mathrm{sw,0}}}\left(\frac{f}{f_{\mathrm{sw}}}\right)^3\left(\frac{7}{4+3\left(\frac{f }{f_{\mathrm{sw}}}\right)^2}\right)^{3.5},\\[5pt]
&h^2\Omega_{\mathrm{tu}}(f)=h^2{\Omega_{\mathrm{tu,0}}}\frac{\left(\frac{f}{f_{\mathrm{tu}}}\right)^2}{\left(1+\frac{f}{f_{\mathrm{tu}}}\right)^{{11}/{3}}\left(1+\frac{8\pi f_{\mathrm{tu}}}{h^\star}\frac{f}{f_{\mathrm{tu}}}\right)},
\end{eqnarray}
where the crest factors are
\begin{eqnarray}
&h^2{\Omega_{\mathrm{en,0}}}=3.6\times 10^{-5}\left(\frac{H_\star}{\beta}\right)^2\left(\frac{\kappa_{\mathrm{en}}\alpha}{1 +\alpha}\right)^2\left(\frac{10}{g^\star}\right)^{\frac{1}{3}}\Gamma,\\[5pt]
&h^2{\Omega_{\mathrm{sw,0}}}=5.7\times 10^{-6}\left(\frac{H_\star}{\beta}\right)\left(\frac{\kappa_{\mathrm{sw}}\alpha}{1+\alpha}\right)^2\left(\frac{10}{g^\star}\right)^{\frac{1}{3}}v_w,\\[5pt]
&h^2{\Omega_{\mathrm{tu,0}}}=7.2\times 10^{-4}\left(\frac{H_\star}{\beta}\right)\left(\frac{\kappa_{\mathrm{tu}}\alpha}{1+\alpha}\right)^2\left(\frac{10}{g^\star}\right)^{\frac{1}{3}}v_w,
\end{eqnarray}
where $\Gamma = {0.11v_w^3}/{(0.42+v_w^2)}$. And the peak frequencies are
\begin{eqnarray}
&\!f_{\mathrm{en}}\!=\!1.12\times \! 10^{-8}\frac{0.62}{1.8-0.1v_w+v_w^2}\left(\frac{\beta}{H_\star}\right)\frac{T_\star}{ 100\mathrm{MeV}}\left(\frac{g^\star}{10}\right)^{\frac{1}{6}},\,\,\,\,\,\,\,\,\\
&f_{\mathrm{sw}}=1.29\times 10^{-8}\frac{1}{v_w}\left(\frac{\beta}{H_\star}\right)\frac{T_\star}{100\mathrm{MeV}}\left(\frac{g ^\star}{10}\right)^{\frac{1}{6}},\\[5pt]
&f_{\mathrm{tu}}=1.84\times 10^{-8}\frac{1}{v_w}\left(\frac{\beta}{H_\star}\right)\frac{T_\star}{100\mathrm{MeV}}\left(\frac{g ^\star}{10}\right)^{\frac{1}{6}},
\end{eqnarray}
where $h$ represents the reduced Hubble parameter, $H_\star$ signifies the Hubble parameter at the transition temperature $T_\star$, $g^\star$ denotes the number of effective degrees of freedom (taking the value of $g^\star = 47.5$ for the PNJL model).

In terms of choosing the characteristic temperature $T_\star$ during the phase transition, we approximately consider the transition temperature as the CEP temperature equals to the critical temperature of chiral phase transition and deconfinement phase transition $T_c=131\,\mathrm{MeV}$. Upon obtaining these crucial parameters, the gravitational wave energy spectrum resulting from the QCD first-order chiral phase transition can be computed.

\begin{figure*}[htpb]
    \centering
    \includegraphics[width=13cm]{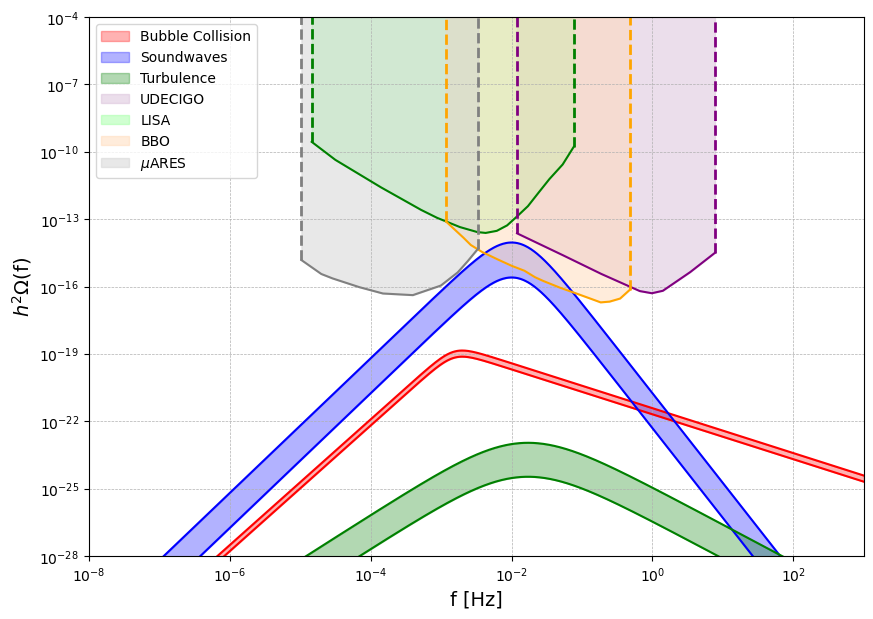}
    \caption{The gravitational waves spectrum of the first-order QCD chiral phase transition. The image is plotted in logarithmic coordinates, where the y-axis represents the gravitational waves spectrum $h^2\Omega (f)$, and the x-axis corresponds to the frequency $f$.}
    \label{Fig.main1}
\end{figure*}

From Figure~\ref{Fig.main1}, we gain an insightful perspective into the dynamic spectrum of gravitational waves generated during the early universe’s phase transitions. The peak frequency of these waves is notably situated around \( f = 10^{-2} \) Hz, which provides a crucial reference point for future observations and theoretical models. Among the various mechanisms contributing to this gravitational wave spectrum, it is evident that sound waves play a pivotal role, particularly in the lower frequency range. Their influence spans a broad frequency band, from \( 10^{-4} \) Hz to 1 Hz, showcasing the diverse impacts of cosmic phenomena on gravitational wave generation. However, this dominance of sound waves in shaping the gravitational wave spectrum is not uniform across all frequencies. As we shift towards higher frequencies, particularly beyond \( f = 1 \) Hz, there is a noticeable decline in the gravitational waves attributed to sound waves. This decrease is so pronounced that at approximately \( f = 1 \) Hz, the gravitational waves generated by bubble collisions surpass those produced by sound waves. Further, at around \( f = 100 \) Hz, the contribution from MHD turbulence becomes more significant than that of the sound waves, though still relatively minor in the overall spectrum.

\section{Conclusions}
\label{sec5}

In this study, we investigate the gravitational waves produced by the QCD phase transition within the purview of the PNJL model. We made an assumption about the adiabatic evolution of the bubbles carrying the chiral symmetry during the phase transition, and concluded that the bubbles gradually runaway during the evolution. Finally, we plotted the gravitational wave spectrum and compared it with existing gravitational wave detection methods.

We reasonably select an adiabatic evolutionary trajectory characterized by an entropy density to energy density ratio of $s/\rho_B=4$ to pinpoint the critical junctures of the phase transition. This approach, which diverges from the conventional assumption of isothermal transitions, augments the robustness of our computational analysis. Furthermore, it furnishes a benchmark for future inquiries focused on delineating the loci pertinent to bubble nucleation and phase transition dynamics. 

Our study also reveals that in first-order QCD chiral phase transitions, if the phase transition intensity isn't strong enough (characterized by a small $\alpha$ value), bubbles originating from the false vacuum fail to gain enough resistance factors throughout their evolution. Consequently, they gradually approach the runaway state as their expansion velocity nears the speed of light. On the other hand, based on the assumption that the entropy to energy density ratio $s/\rho_B$ remains constant, when the phase transition intensity is overly strong (symbolized by a large $\alpha$ value), bubbles are unable to reach the true vacuum state through the adiabatic evolution pathway, for the reason that the overly intense phase transition leads to extremely low temperatures and densities after the evolution along the adiabatic pathway, rendering it physically unfeasible.

The results of our investigation highlight a nuanced aspect of cosmological phase transitions: an exceedingly vigorous phase transition may impede the system's progression to the true vacuum state, thereby indicating a diminished probability of such high-intensity phase transitions occurring within the early universe. Since the amplitude and detectability of gravitational waves are intrinsically tied to the intensity of the underlying phase transition, gravitational waves emanating from these infrequent yet intense phase transitions represent a challenging detection frontier. Crucially, our analysis delineates an upper bound on the intensity of chiral phase transitions, a constraint that inherently limits the peak amplitude of the resultant gravitational waves to approximately $10^{-14}$. This finding aligns the peak frequency within a range of $10^{-4}\,\mathrm{Hz}$ to $1\,\mathrm{Hz}$, a spectrum that falls within the anticipated sensitivity of emerging gravitational wave observatories.

Notably, our analysis indicates that the overall intensity of the chiral phase transition, while significant, is not sufficient to produce gravitational waves strong enough for detection by current observatories like LISA. This underscores the challenge in observing these subtle cosmic phenomena and the need for more sensitive detection technology. However, advanced detectors, such as the proposed Big Bang Observer (BBO), are anticipated to have the capability to capture these elusive signals. The potential of these future observatories to detect gravitational waves from phase transitions, especially those occurring in the centihertz frequency range. This revelation steers the focus of gravitational wave detection technology towards this specific frequency domain, thereby enhancing the probability of discerning these subtle cosmic signals.

\section*{Acknowledgements}
This work is supported by the National Natural Science Foundation of China under Grant No.~11875213.

\end{document}